\documentclass[aps,showpacs,preprint,preprintnumbers,amsmath,amssymb,nofootinbib,contrib]{revtex4}
\usepackage{amssymb}
\usepackage{dcolumn}
\usepackage{bm}
\allowdisplaybreaks[2]

\def\beq{\begin{equation}}
\def\eeq{\end{equation}}

\newcommand{\bea}{\begin{eqnarray}}
\newcommand{\eea}{\end{eqnarray}}

\def\gev{\rm GeV}

\def\eeqn{\end{equation}}
\newcommand\iden{\leavevmode\hbox{\small1\normalsize\kern-.33em1}}

\let\jnfont=\rm
\def\NPB#1,{{\jnfont Nucl.\ Phys. {\bf B}}{\bf #1},}
\def\PLB#1,{{\jnfont Phys.\ Lett. {\bf B}}{\bf #1},}
\def\PRD#1,{{\jnfont Phys.\ Rev. {\bf D}}{\bf #1},}
\def\PRL#1,{{\jnfont Phys.\ Rev.\ Lett.\ }{\bf #1},}
\def\ZPC#1,{{\jnfont Z.\ Phys. {\bf C}}{\bf #1},}
\def\MPLA#1,{{\jnfont Mod.\ Phys.\ Lett.\ A }{\bf #1},}
\def\JPG#1,{{\jnfont J.\ Phys.\ G }{\bf #1},}
\def\CTP#1,{{\jnfont Commun.\ Theor.\ Phys.\ }{\bf #1},}
\def\JHEP#1,{{\jnfont JHEP \ }{\bf #1},}
\def\NPPS#1,{{\jnfont Nucl.\ Phys.\ Proc.\ Suppl.\ }{\bf #1},}

\begin{document}


\title{Baryonic Isgur-Wise Functions in Large $N_c$ HQET}

\author{Ming-Kai Du and Chun Liu}
\affiliation{State Key Laboratory of Theoretical Physics, \\
Institute of Theoretical Physics, Chinese Academy of Sciences, \\
P.O. Box 2735, Beijing 100190, China}
\email{mkdu@itp.ac.cn,liuc@mail.itp.ac.cn}

\begin{abstract}
Large $N_c$ relations among baryonic Isgur-Wise functions appearing at
the order of $1/m_Q$ are analyzed.  An application to
$\Omega_b \to \Omega_c$ weak decays is given.
\end{abstract}

\pacs{12.39.Hg, 11.15.Pg, 14.20.Mr, 13.30.Ce}

\keywords{HQET, Large $N_c$, bottom baryons}
\maketitle

\section{Introduction}

Weak decays of heavy baryons are interesting both experimentally and
theoretically.  They are now under the study of the LHC experiments,
as well as previous Tevatron and LEP experiments. They also provide
a playing ground for nonperturbative QCD methods.  Heavy baryons
containing a single heavy quark are described by the heavy quark
effective theory (HQET) \cite{HQET1,HQET2,HQET3,HQET4}.  Relevant
physical quantities can be factorized into a calculable perturbative
part and universal hadronic quantities.  To calculate the latter,
some nonperturbative methods, like the large $N_c$ one \cite{largen1},
are needed.

Consider the heavy baryon weak transitions
$\Lambda_b\rightarrow \Lambda_c$ and
$\Sigma^{(*)}_b\rightarrow \Sigma^{(*)}_c$.
The matrix elements of vector and axial currents
($V^\mu=\bar c\gamma^\mu b$ and
$A^\mu=\bar c\gamma^\mu \gamma^5 b$) between the $\Lambda_b$ and
$\Lambda_c$ can be parametrized as
\bea
\langle\Lambda_c(v',s')|V^\mu|\Lambda_b(v,s)\rangle&=&
\bar u_{\Lambda_c}(v',s')(F_1(\omega)\gamma^{\mu}+F_2(\omega)v^{\mu}
+F_3(\omega)v'^{\mu})u_{\Lambda_b}(v,s),
\nonumber\\
\langle\Lambda_c(v',s')|A^\mu|\Lambda_b(v,s)\rangle&=&
\bar u_{\Lambda_c}(v',s')(G_1(\omega)\gamma^{\mu}+G_2(\omega)v^{\mu}
+G_3(\omega)v'^{\mu})\gamma^5 u_{\Lambda_b}(v,s),
\label {eq:general lambda}
\eea
and those between $\Sigma_b$ and $\Sigma^{(*)}_c$ are
\bea
\langle\Sigma_c(v',s')|V^\mu|\Sigma_b(v,s)\rangle&=&
\bar u_{\Sigma_c}(v',s')(F'_1\gamma^{\mu}+F'_2v^{\mu}+F'_3v'^{\mu})u_{\Sigma_b}(v,s),
\nonumber\\
\langle\Sigma_c(v',s')|A^\mu|\Sigma_b(v,s)\rangle&=&
\bar u_{\Sigma_c}(v',s')(G'_1\gamma^{\mu}+G'_2v^{\mu}+G'_3v'^{\mu})\gamma^5 u_{\Sigma_b}(v,s),
\nonumber \\
\langle\Sigma^*_c(v',s')|V^\mu|\Sigma_b(v,s)\rangle&=&
\bar u_{\Sigma^*_c\lambda}(v',s')(N'_1v^{\lambda}\gamma^{\mu}
+N'_2v^{\lambda}v^{\mu}+N'_3v^{\lambda}v'^{\mu}
+N'_4g^{\lambda\mu})\gamma^5u_{\Sigma_b}(v,s),
\nonumber \\
\langle\Sigma^*_c(v',s')|A^\mu|\Sigma_b(v,s)\rangle&=&
\bar u_{\Sigma^*_c\lambda}(v',s')(K'_1v^{\lambda}\gamma^{\mu}
+K'_2v^{\lambda}v^{\mu}+K'_3v^{\lambda}v'^{\mu}+K'_4g^{\lambda\mu})u_{\Sigma_b}(v,s),
\nonumber \\
&&
\label {eq:general sigma}
\eea
where $\omega=v\cdot v'$ and $F_i(\omega)^{(\prime)}$,
$G_i(\omega)^{(\prime)}$, $N_i(\omega)'$ and $K_i(\omega)'$ are general
form factors.  As is well known, in the HQET, form factors can be
described in terms of several independent universal form factors which
are the so-called Isgur-Wise functions.

The Isgur-Wise functions are defined as follows.  Note that it is
self-evident that in the HQET, heavy quark fields and baryon fields
have their own definition, in spite of adopting the same symbols as in full QCD.
For the $\Lambda_b\to\Lambda_c$ transition, at the leading order of heavy
quark expansion, there is only one Isgur-Wise function $\eta(\omega)$,
\beq
\langle\Lambda_c(v',s')|\bar c\Gamma b|\Lambda_b(v,s)\rangle=
\eta(\omega)\bar u_{\Lambda_c}(v',s')\Gamma u_{\Lambda_b}(v,s) \,,
\label{eq:lambda}
\eeq
where $\Gamma$ stands for general $\gamma$ matrices, and
$\eta(\omega)$ is normalized at the zero recoil, namely $\eta(1)=1$.
For $\Sigma^{(*)}_b\to \Sigma^{(*)}_c$ transitions, two Isgur-Wise
functions $\xi_1(\omega)$ and $\xi_2(\omega)$ appear
at the leading order \cite{leading order},
\beq
\langle\Sigma^{(*)}_c(v',s')|\bar c\Gamma b|\Sigma^{(*)}_b(v,s)\rangle
=\left[-g_{\mu\nu}\xi_1(\omega)+v_\mu v'_\nu \xi_2(\omega)\right]
\bar u^\mu_{\Sigma_c^{(*)}}(v',s')\Gamma
u^\nu_{\Sigma_b^{(*)}}(v,s) \,,
\label{eq:sigma1}
\eeq
where $\xi_1(1)=1$, and $u^\nu_{\Sigma_Q^*}$ is the Rarita-Schwinger
spinor for a spin-$\frac32$ particle.  And $u^\nu_{\Sigma_Q}$ is defined
as
\beq
u^\nu_{\Sigma_Q}(v,s)=
\frac{\gamma^\nu+v^\nu}{\sqrt3}\gamma_5u_{\Sigma_Q}(v,s) \,.
\eeq
Isgur-Wise functions at $1/m_Q$ order will be discussed in detail in
the next section.

Then, in the heavy quark limit, the general form factors in
Eqs. (\ref{eq:general lambda}) and (\ref{eq:general sigma}) are
simplified.  For the $\Lambda_b\to\Lambda_c$ transition,
\beq
F_1=G_1=C(\mu)\eta(\omega)\qquad,\qquad F_2=G_2=F_3=G_3=0\, ,
\eeq
where $C(\mu)$ is a perturbatively calculable coefficient. For
$\Sigma^{(*)}_b\to \Sigma^{(*)}_c$ transitions, the formulas are
a bit more complex, which can be found in \cite{boyd brahm}.  Note that
Isgur-Wise functions are independent of the weak currents.

At this stage, the Isgur-Wise functions are still unknown, and need
nonperturbative methods to be calculated. In the large $N_c$ limit,
interesting information about baryonic Isgur-Wise functions was
obtained.  In large $N_c$ baryons, there is a spin-flavor symmetry
of light quarks \cite{largen2} which not only gives the mass
degeneracy of $\Sigma^{(*)}_Q$ and $\Lambda_Q$, but also results in
the following relations among the Isgur-Wise functions
\cite {Universal1,Universal2}:
\beq
\xi_1(\omega)  =  \eta(\omega) \,,\quad
\xi_2(\omega)  =  \frac{\eta(\omega)}{1+\omega} \,.
\label{eq:leading_large_n}
\eeq
This large $N_c$ result is also consistent with that obtained from
the large $N_c$ constituent quark model \cite{quarkmodel}.

Furthermore, in the heavy baryon Skyrme model \cite{skyrme},
$\eta(\omega)$ is calculated to be \cite{largen3}
\beq
\eta(\omega)=0.99\, \textrm{exp}[-1.3(\omega-1)].
\label{eq:IW}
\eeq
In fact, in the real large $N_c$ limit, $\eta(\omega)$ is actually a
$\delta$-function \cite{largen4}.

From Eqs. (\ref{eq:lambda}), (\ref{eq:sigma1}) and
(\ref{eq:leading_large_n}), it is observed that to obtain the
$\Sigma^{(*)}_Q$ matrix elements of weak currents in large $N_c$
approximation, what we need to do is just multiplying the $\Lambda_Q$
matrix element by the following Lorentz tensor:
\beq
\bigg[-g_{\mu\nu}+ \frac {v_\mu v'_\nu}{1+\omega}\bigg] \,.
\eeq
This is because the two kinds of decays are essentially the same, except
for Lorentz structures (a kinetic result of the light degrees of
freedom).

With this observation, in the following sections, we will extend the
large $N_c$ relations in Eq. (\ref{eq:leading_large_n}) to $O(1/m_Q)$.

\section{Isgur-Wise functions at $O(1/m_Q)$}

There are more Isgur-Wise functions at the $1/m_Q$ order.  Before we
consider large $N_c$ relations among the Isgur-Wise functions at
$O(1/m_Q)$, it is useful to start from their definition.

\subsection{A mini-review}

$1/m_Q$ corrections arise from two sources.  One is due to the HQET
Lagrangian at $O(1/m_Q)$, and the other is obtained through the
$1/m_Q$ expansion of heavy quark currents in the full QCD.

Firstly, let us consider the Lagrangian corrections.  To the order
$O(1/m_Q)$ the effective Lagrangian is \cite {HQET2,luke}
\bea
\mathcal L&=&\bar h_v iv\cdot D h_v
\nonumber \\
&&-\frac 1{2m_Q}\bar h_v[D^2
+\frac 12g_s\sigma_{\mu\nu}G^{\mu\nu}]h_v \,.
\nonumber
\eea
For the hadronic matrix element of a heavy quark current, correction
due to the heavy quark kinetic energy is
\beq
\langle H_c| -i\int d^4xT\bigg(g_s\bar c_{v'}\frac{D^2}{2m_c}
c_{v'}\bigg|_x\bar c_{v'}\Gamma b_v\bigg|_0\bigg)|H_b\rangle \,.\\
\label{eq:ckinetic}
\eeq
For $H_Q$ being $\Lambda_Q$, Eq. (\ref{eq:ckinetic}) is parametrized
as
\beq
\bar u_{\Lambda_c}(v',s')\frac{1+v'\!\!\!\!/}2
\Gamma u_{\Lambda_b}(v,s)\frac {\chi(\omega)}{m_c} \,,
\label{eq:chidef}
\eeq
where $\chi(\omega)$ is the $\Lambda_Q$ Isgur-Wise function at the order
of $1/m_Q$.  It satisfies that $\chi(1)=0$.  In the real large $N_c$ limit
$\chi(\omega)=0$ \cite{largen4}.

In the case of $H_Q$ being $\Sigma^{(*)}_Q$, Eq. (\ref{eq:ckinetic})
is parametrized via two more Isgur-Wise functions,
\beq
\frac 1 {m_c}\Big[-g_{\mu\nu}\chi_1(\omega)
+v_\mu v'_\nu \chi_2(\omega)\Big]\bar u^\mu_{\Sigma_c^{(*)}}(v',s')
\frac{1+v'\!\!\!\!/}2\Gamma u^\nu_{\Sigma_b^{(*)}}(v,s)\, ,
\label{14}
\eeq
with that $\chi_1(1)=\chi_2(1)=0$.

Correction due to the heavy quark chromomagnetic interaction is
\beq
\langle H_Q|-\frac i2 \int d^4x T \bigg ( g_s\bar c_{v'}
\frac {\sigma_{\mu\nu}G^{\mu\nu}}{2m_c}c_{v'}\bigg|_x
\bar c_{v'}\Gamma b_v\bigg|_0 \bigg)|H_Q\rangle \,.
\eeq
For $H_Q$ being $\Lambda_b$, it is zero \cite{HQET3}.  It is a bit
more complicated in the case of $H_Q$ being $\Sigma^{(*)}_Q$.
According to Ref. \cite{boyd brahm}, the chromomagnetic $1/m_Q$
correction can be parametrized as
\beq
\bar u^\mu_{\Sigma_c^{(*)}}(v',s')\sigma^{\lambda\rho}
\frac{1+v'\!\!\!\!/}2\Gamma
u^\nu_{\Sigma_b^{(*)}}(v,s)M^{\mu\nu}_{\lambda\rho},
\eeq
where
\beq
M^{\mu\nu}_{\lambda\rho}=
\frac1 {2m_c} \Big[\zeta_1(\omega)g^\mu_\lambda g^\nu_\rho
+\zeta_2(\omega)g^\mu_\rho v'^\nu v_\lambda
+\zeta_3(\omega)g^\nu_\lambda v^\mu v_\rho\Big] .
\eeq

Now consider $1/m_Q$ corrections of the current operator.  The
relation between the QCD currents and HQET operators is
\beq
\bar c \Gamma b = \bar c_{v'}\bigg(\Gamma
-\frac{i\overleftarrow D_\alpha}{2m_Q}
\gamma^\alpha\Gamma\bigg) b_v \quad \,.
\eeq
For the $\Lambda_b\to\Lambda_c$ decay \cite {luke,1/m},
\beq
\bar c_{v'}i\overleftarrow D_\alpha\Gamma b_v=
\frac{\bar\Lambda\eta}{(1+\omega)}\bar u_{\Lambda_c}(v',s')\Gamma
u_{\Lambda_b}(v,s)(v_\alpha-\omega v'_\alpha) \quad ,
\eeq
where $\bar\Lambda= m_{\Lambda_Q}-m_Q$.
For the $\Sigma^{(*)}_b(v,s)\to\Sigma^{(*)}_c(v',s')$ transition, the
$1/m_Q$ correction is parametrized as \cite{boyd brahm}
\beq
\bar c_{v'}i\overleftarrow D_\alpha\Gamma b_v=
\bar u^\mu_{\Sigma_c^{(*)}}(v',s')\Gamma
u^\nu_{\Sigma_b^{(*)}}(v,s)P^{\mu\nu}_\alpha,
\eeq
where
\beq
P^{\mu\nu}_\alpha=\kappa_1v'^\nu v^\mu v_\alpha
+\kappa_2v'^\nu v^\mu v'_\alpha+\kappa_3g^{\mu\nu}v_\alpha
+\kappa_4g^{\mu\nu}v'_\alpha+\kappa_5g^\mu_\alpha v'^\nu
+\kappa_6g^\nu_\alpha v^\mu.
\eeq
Actually, only two of these Isgur-Wise functions are independent
\cite {boyd brahm},
\beq
\kappa_3=-\frac{\bar\Sigma}{1+\omega}\xi_1 ,\quad
\kappa_4=\frac{\bar\Sigma\omega}{1+\omega}\xi_1 , \quad
\kappa_5=\bar\Sigma(1-\omega)\xi_2-(\kappa_1+\omega\kappa_2), \quad
\kappa_6=-(\omega\kappa_1+\kappa_2).
\label {eq:relations}
\eeq

The expressions of form factors in
Eqs. (\ref{eq:general lambda}) and (\ref{eq:general sigma}) in terms of
all these Isgur-Wise functions can be found in \cite{boyd brahm}.

\subsection{Large $N_c$ relations}

Now consider the large $N_c$ limit, there are relations among the
subleading order Isgur-Wise functions.  The point is that the
observation in Sec. I is still applicable.  The only difference here is
that heavy quark currents have different forms, which are irrelevant
because of the heavy quark symmetry.  Then in the large $N_c$ limit,
Eq. (\ref{14}) which is the charm quark kinetic energy correction should
be
\beq
\frac {\chi(\omega)}{m_c}\Big[-g_{\mu\nu}+
\frac{v_\mu v'_\nu}{(1+\omega)}\Big]
\bar u^\mu_{\Sigma_c^{(*)}}(v',s')\Gamma u^\nu_{\Sigma_b^{(*)}}(v,s)\,.
\eeq
This results in the following relations:
\beq
\chi_1(\omega)  =  \chi(\omega) ,\quad
\chi_2(\omega)  =  \frac {\chi(\omega)}{1+\omega}\,.
\label {eq:universal1}
\eeq

With the same method, we obtain a pleasant result: in the large $N_c$
limit, to the $1/m_Q$ order in HQET, there is no chromomagnetic
corrections in the $\Sigma^{(*)}_b(v,s)\rightarrow\Sigma^{(*)}_c(v',s')$
decay,
\beq
\zeta_1(\omega)=\zeta_2(\omega)=\zeta_3(\omega)=0.
\label {eq:universal2}
\eeq
This can be understood easily, since in the large $N_c$ limit, spins and
isospins of light degrees of  freedom in $\Sigma^{(*)}_b(v,s)$ and
$\Sigma^{(*)}_c(v',s')$ have decoupled.  This decoupling makes
$\Sigma^{(*)}_Q$ no different from $\Lambda_Q$.

Therefore, in the large $N_c$ limit, the time-ordered product of $1/m_Q$
terms in the Lagrangian with the heavy quark current just produces a
trivial correction for $\Sigma^{(*)}_Q$ decays:
a redefinition of the leading order Isgur-Wise
functions, this is similar to the case of $\Lambda_Q$ decays.

Looking at the $1/m_c$ correction of the current operator, for
$\Sigma^{(*)}_Q$ we have
\beq
\bar c_{v'}i\overleftarrow D_\alpha\Gamma b_v=
\frac{\bar\Sigma\eta}{(1+\omega)}(v_\alpha-\omega v'_\alpha)
\Big[-g_{\mu\nu}+\frac{v_\mu v'_\nu}{1+\omega}\Big]
\bar u^\mu_{\Sigma_c^{(*)}}(v',s')\Gamma u^\nu_{\Sigma_b^{(*)}}(v,s),
\eeq
where $\bar\Sigma$ is defined as
$\bar\Sigma=m_{\Sigma_b}-m_b\simeq m_{\Sigma_c}-m_c$. Note that
$\bar\Sigma=\bar\Lambda$ in the large $N_c$ limit.
Again, we obtain some new relations as below:
\beq
\kappa_1 =  \frac{\bar\Sigma}{(1+\omega)^2}\eta,\quad
\kappa_2 = -\frac{\bar\Sigma\omega}{(1+\omega)^2}\eta,\quad
\kappa_3 = -\frac{\bar\Sigma}{1+\omega}\eta,\quad
\kappa_4 =  \frac{\bar\Sigma\omega}{1+\omega}\eta,\quad
\kappa_5 = \kappa_6 =0.
\label {eq:universal3}
\eeq
It is observed that, after taking large $N_c$ approximation, the
relations obtained in HQET, such as Eq.(\ref{eq:relations}),
still hold.

\subsection{General form factors}

Up to now, we have derived all of large $N_c$ relations for the $1/m_c$
corrections, $1/m_b$ corrections can be obtained similarly.
Including all $1/m_Q$ corrections, in the large $N_c$ limit, the general
form factors in Eqs. (\ref{eq:general lambda}) and (\ref{eq:general sigma}) are
expressed as
\bea
&&F_1=\eta'(\omega)+\eta'(\omega)\Bigg[\frac{\bar\Lambda}{2m_c}+
\frac{\bar\Lambda}{2m_b}\Bigg]\ ,\qquad\qquad\quad
  G_1=\eta'(\omega)-\eta'(\omega)\Bigg[\frac{\bar\Lambda}{2m_c}+
\frac{\bar\Lambda}{2m_b}\Bigg]\Bigg(\frac{1-\omega}{1+\omega}\Bigg)
\nonumber \\[2mm]
&&F_2=-\frac{\bar\Lambda}{m_c}\Bigg(\frac1{1+\omega}\Bigg)\eta'(\omega)
\ ,\qquad\qquad\quad\quad\;\qquad
  G_2=-\frac{\bar\Lambda}{m_c}\Bigg(\frac1{1+\omega}\Bigg)\eta'(\omega)
\nonumber \\[2mm]
&&F_3=-\frac{\bar\Lambda}{m_b}\Bigg(\frac1{1+\omega}\Bigg)\eta'(\omega)
\ ,\qquad\qquad\qquad\quad\;\quad
  G_3= \frac{\bar\Lambda}{m_b}\Bigg(\frac1{1+\omega}\Bigg)\eta'(\omega)
\nonumber \\[2mm]
&&F'_1=-\frac13 \eta'(\omega)-\frac13 \eta'(\omega)\Bigg[\frac{\bar\Sigma}{2m_c}+
\frac{\bar\Sigma}{2m_b}\Bigg]
\ ,\qquad\quad
  G'_1=-\frac13 \eta'(\omega)+\frac13 \eta'(\omega)\Bigg[\frac{\bar\Sigma}{2m_c}+
\frac{\bar\Sigma}{2m_b}\Bigg]\Bigg(\frac{1-\omega}{1+\omega}\Bigg)
\nonumber \\[2mm]
&&F'_2= \frac{4\eta'(\omega)}{3(1+\omega)}+\frac{\eta'(\omega)}{3(1+\omega)}
\Bigg[-\frac{\bar\Sigma}{m_c}+\frac{2\bar\Sigma}{m_b}\Bigg]
\ ,\qquad
G'_2=\frac{\bar\Sigma}{3m_c}\Bigg(\frac1{1+\omega}\Bigg)\eta'(\omega)
\nonumber\\ [2mm]
&&F'_3= \frac{4\eta'(\omega)}{3(1+\omega)}+\frac{\eta'(\omega)}{3(1+\omega)}
\Bigg[\frac{2\bar\Sigma}{m_c}-\frac{\bar\Sigma}{m_b}\Bigg]
\ ,\quad\qquad
G'_3=-\frac{\bar\Sigma}{3m_b}\Bigg(\frac1{1+\omega}\Bigg)\eta'(\omega)
\nonumber\\ [2mm]
&&N'_1=\frac{-2\eta'(\omega)}{\sqrt3\Big(1+\omega\Big)}+\frac{-\eta'(\omega)}{\sqrt3\Big(1+\omega\Big)}
\Bigg[\frac{\bar\Sigma}{m_c}+\frac{\bar\Sigma}{m_b}\Bigg]
\ ,\
  K'_1=0
\nonumber \\ [2mm]
&&N'_2=0\ ,\qquad
K'_2=\frac2{\sqrt3}{\eta'(\omega)}\frac{\bar\Sigma}{m_c}
\Bigg(\frac1{1+\omega}\Bigg)^2\ ,
\nonumber \\ [2mm]
&&N'_3=\frac{2\eta'(\omega)}{\sqrt3\Big(1+\omega\Big)}+\frac{\eta'(\omega)}{\sqrt3\Big(1+\omega\Big)}
\Bigg[\frac{\bar\Sigma}{m_c}+\frac{\bar\Sigma}{m_b}\Bigg]
\ ,\quad \nonumber \\ [2mm]
&&K'_3=\frac{-2\eta'(\omega)}{\sqrt3\Big(1+\omega\Big)}+\frac{\eta'(\omega)}{\sqrt3\Big(1+\omega\Big)}
\Bigg[\frac{\bar\Sigma}{m_c}\Bigg(\frac{1-\omega}{1+\omega}\Bigg)
-\frac{\bar\Sigma}{m_b}\Bigg],
\nonumber \\ [2mm]
&&N'_4=\frac{-2\eta'(\omega)}{\sqrt3}-\frac1{\sqrt3} \eta'(\omega)
\Bigg[\frac{\bar\Sigma}{m_c}+\frac{\bar\Sigma}{m_b}\Bigg],
\nonumber \\ [2mm]
&&K'_4=\frac{2\eta'(\omega)}{\sqrt3}-\frac1{\sqrt3} \eta'(\omega)
\Bigg[\frac{\bar\Sigma}{m_c}+\frac{\bar\Sigma}{m_b}\Bigg]
\Bigg(\frac{1-\omega}{1+\omega}\Bigg),
\label{eq:results}
\eea
where
\beq
\eta'(\omega)\equiv \eta(\omega)
+\chi(\omega)\bigg(\frac 1 {m_c}+\frac 1 {m_b}\bigg)\quad ,
\eeq
and all the form factors should be multiplied by $C(\mu)$ in
Eq.(\ref{eq:lambda}).  We have checked that all the results are
consistent with \cite{boyd brahm} where all $1/m_Q$ form factors are
listed.  After taking the large $N_c$ limit, all the relations of
$1/m_Q$ form factors shown in \cite{boyd brahm} still hold, especially
the following normalization relations at zero recoil point:
\bea
F_1(1)+F_2(1)+F_3(1)=C(\mu)\ ,      &\quad& G_1(1)=C(\mu)\ ;\nonumber \\
F'_1(1)+F'_2(1)+F'_3(1)=C(\mu) \ ,  &\quad& G'_1(1)=-\frac 13C(\mu)\ ,\quad
K'_4(1)=\frac 2{\sqrt3}C(\mu) \quad .
\eea

In fact, through our analysis, it is easy to see that the large $N_c$
limit and HQET are commutative, in other words, the large $N_c$
approximation preserves all relations obtained in HQET.

\section{The weak decays}

As an application, we now calculate
$\Omega_b \rightarrow \Omega_{c}^{(*)}$ weak decay rates \cite{duliu}.
Since $\Sigma^{(*)}_b$ has the strong interaction decay mode, we
mainly consider the semileptonic decays of $\Omega_b$.  In the SU(3)
light quark flavor symmetry limit, $\Omega_{b(c)}^{(*)}$ baryons
are identical to $\Sigma_{b(c)}^{(*)}$ baryons.  Therefore, for the
Isgur-Wise functions, the same results for $\Omega_{b(c)}^{(*)}$ can be
obtained.

Neglecting the lepton masses, for the decay of
$\Omega_b\rightarrow\Omega_c\ l\ \bar\nu$, the differential decay rate
can be expressed \cite{decay rate1,decay rate2} in terms of the general
form factors in Eq. (\ref{eq:general lambda}) as
\bea
\frac{d\Gamma_1(\omega)}{d\omega}= && \frac{G^2_F|V_{cb}|^2m^5_{\Omega_b}r^3_2}
{24\pi^3}\sqrt{(\omega^2-1)}\nonumber\\
&&\times\bigg\{2(\omega-1)\kappa_2 F'^2_1+(\omega-1)\Big[(1+r_2)F'_1+
(\omega+1)\Big(r_2F'_2+F'_3\Big)\Big]^2\nonumber \\
&&+2(\omega+1)\kappa_2 G'^2_1+(\omega+1)\Big[(1-r_2)G'_1-
(\omega-1)\Big(r_2G'_2+G'_3\Big)\Big]^2\bigg\}
\eea
where $r_2=m_{\Omega_c}/m_{\Omega_b}$ and $\kappa_2=1+r^2_2-2r_2\omega$.

For the decay of $\Omega_b\rightarrow\Omega^*_c\ l\ \bar\nu$, we
have \cite{decay rate3}
\bea
\frac{d\Gamma_2(\omega)}{d\omega}= && \frac{G^2_F|V_{cb}|^2m^5_{\Omega_b}r^3_3}
{72\pi^3}\sqrt{(\omega^2-1)}\nonumber\\
&&\times\Bigg\{(\omega-1)\kappa_3\Big[N'_4-2(\omega+1)N'_1\Big]^2+
        (\omega+1)\kappa_3\Big[K'_4-2(\omega-1)K'_1\Big]^2
        \nonumber\\
&& +2(\omega+1)\Big[(\omega-1)(r_3+1)K'_1+(\omega^2-1)(K'_3+r_3K'_2)+
(\omega-r_3)K'_4\Big]^2\nonumber\\
&& +2(\omega-1)\Big[(\omega+1)(r_3-1)N'_1+(\omega^2-1)(N'_3+r_3N'_2)+
(\omega-r_3)N'_4\Big]^2\nonumber\\
&&+3\kappa_3\Big[(\omega+1)K'^2_4+(\omega-1)N'^2_4\Big]
\Bigg\}
\eea
where $r_3=m_{\Omega^*_c}/m_{\Omega_b}$ and
$\kappa_3=1+r^2_3-2r_3\omega$.

The form factors have been expanded in Eq.(\ref{eq:results}) to the order
of $1/m_c$ and $1/m_b$.  There are only one Isgur-Wise function
$\eta(\omega)$ at the leading order, one $\chi(\omega)$ at the
subleading order, and the mass parameter $\bar\Omega\equiv\bar\Sigma$.
With Eq. (\ref{eq:IW}) and $\chi(\omega) \simeq 0 $ \cite{largen4}, we
obtain the decay widths as
\bea
\Gamma(\Omega_b\rightarrow\Omega_c\ l\ \bar\nu) = 3.38\times10^{-14}\ \gev;\nonumber\\
\Gamma(\Omega_b\rightarrow\Omega^*_c\ l\ \bar\nu) = 3.34\times10^{-14}\ \gev.
\eea
In the calculations, we have taken the following parameters \cite{pdg}:
\bea
m_{\Omega_b} = 6.07\ \textrm{Gev}&\quad,\quad&
m_{\Omega_c} = 2.70\ \textrm{Gev},\quad\nonumber \\
m_{\Omega^*_c} = 2.77\ \textrm{Gev}&\quad,\quad&
\mid V_{cb}\mid = 40.9\times10^{-3}\ .
\eea
The pole masses of heavy quarks have been taken as
$m_b = 4.83\ \textrm{Gev}, \ m_c = 1.43\ \textrm{Gev}$.

\section{Discussions}

In this paper, we have studied $O(1/m_Q)$ universal baryonic Isgur-Wise
functions in the large $N_c$ limit, our results are explicitly listed in
Eqs. (\ref{eq:universal1}), (\ref{eq:universal2}) and (\ref{eq:universal3}).  
As an application, we have calculated the
semileptonic decays of $\Omega_b$.
Actually, the same results would be obtained using the leading order
large $N_c$ analysis in \cite {Universal1,Universal2,quarkmodel}, while
our method is a lot simpler.

Let us now consider the uncertainties of our results.  Since the $1/m_Q$
corrections have been included, the uncertainties brought about by HQET
have been suppressed to \textit{O}($\Lambda_{QCD}^2/m_c^2\sim1/25$). The
remaining uncertainties come from two approximations: flavor SU(3)
symmetry and large $N_c$ limit.  Effects of flavor SU(3) violation might
not be huge especially near the zero recoil point. Since $\Omega_b$ has
s-quarks, we could expect the effects as \cite{HQET3}:
\beq
\Big|\xi_1(\omega)-\eta(\omega)\Big|\sim \ln\Bigg(\frac {m^2_K} {\mu^2}\Bigg)
\eeq
which would be small by choosing some appropriate renormalization scale
$\mu$.  Then, the main uncertainties are produced by the large $N_c$
approximation, while we have so little knowledge about them.  Sometimes
they can be as large as $30\%$. In our case, however, it is unnecessary to be that
pessimistic.  As a general experience, the large $N$ limit is a good
approximation for baryons and also good for Isgur-Wise functions.
Because of replacing $\bar\Sigma$ with $\bar\Omega$ in the decay
calculation, we have already taken part of the flavor SU(3) violation
effects and part of the corrections to the large $N_c$ limit into account.

Finally, it is important to notice that, whether or not the large $N_c$
limit can be treated as a good approximation, at least in the vicinity
of the zero recoil point, the uncertainties produced by large $N_c$
limit should not be huge, since it preserves the normalizations of
Isgur-Wise functions as in HQET, just like we have emphasized in Sec.III,
which will be tested at the LHC or the proposed Z-factory in the near future.

\acknowledgments

This work was supported in part by the National Natural Science
Foundation of China under nos. 11075193 and 10821504, and by the
National Basic Research Program of China under Grant No. 2010CB833000.

\appendix*


\end{document}